\documentstyle[twocolumn,prb,aps]{revtex}

\def\etal{{\sl et al\/}\ }

\draft
\begin{document}

\title{Quasi Harmonic Lattice Dynamics and Molecular Dynamics calculations \\
for the Lennard-Jones solids}

\author{Raffaele Guido Della Valle and Elisabetta Venuti}

\address{Dipartimento di Chimica Fisica e Inorganica,
Universit\`a di Bologna, Viale Risorgimento 4, I-40136 Bologna, Italy}


\date{\today}
\wideabs{
\maketitle
\begin{abstract}
We present Molecular Dynamics (MD), Quasi Harmonic Lattice Dynamics
(QHLD) and Energy Minimization (EM) calculations for the crystal
structure of Ne, Ar, Kr and Xe as a function of pressure and
temperature. New Lennard-Jones (LJ) parameters are obtained for Ne, Kr
and Xe to reproduce the experimental pressure dependence of the
density. We employ a simple method which combines results of QHLD and
MD calculations to achieve densities in good agreement with experiment
from 0 K to melting. Melting is discussed in connection with
intrinsic instability of the solid as given by the QHLD approximation.
\end{abstract}
}

\section{Introduction}

Lattice Dynamics (LD) is based on the expansion of the potential energy of
the crystal in powers of the displacements of the atoms from a reference
structure \protect\cite{Born1954,Pertsin1987}. The harmonic approximation, in which the
expansion is truncated at the second order, reduces the many-body problem
to many exactly soluble one-body problems and allows a direct computation
of all thermodynamic functions. Harmonic \protect\cite{Born1954} and quasi-harmonic \protect\cite{Ludwig1967,Zharkov1971,Barron1974} lattice dynamics (HLD and QHLD) are
alternative LD strategies, which differ in the choice of the reference
structure. For HLD this is the minimum potential energy (mechanical
equilibrium) structure, whereas for QHLD it is the minimum free energy
(thermodynamic equilibrium) average structure at a given temperature $T$
and pressure $p$. This difference is crucial, because the accuracy of the
harmonic approximation is controlled by the amplitude of the displacements
from the reference structure. Therefore QHLD might remain accurate in
regimes where the average structure deviates significantly from that of
the potential minimum, provided that the amplitude of the vibrations
around the average does not grow too large. Conventional HLD is definitely
unusable in these situations.

A straightforward way to assess the accuracy of the QHLD approximation is
to choose a model system and to compare the QHLD calculations with the
essentially ``exact'' classical mechanics results of Molecular Dynamics
(MD) or Monte Carlo (MC) simulations \protect\cite{Allen1987}. If the same model system
and interaction potential are employed for both QHLD and the classical
simulations, all differences between the two sets of results must be due to
the approximations implicit in the two methods. 

Lacks and Rutledge \protect\cite{Lacks1994} have chosen a monoatomic {\it fcc} lattice with
Lennard-Jones (LJ) interactions as a model to study the behaviour of the
QHLD and MC free energies as a function of $T$. They find that the QHLD free
energy is reasonably accurate up to $T_m/2$, and rapidly deteriorates above
that; here $T_m$ is the melting temperature. These results clearly indicate
that QHLD as such is not reliable close to $T_m$. However, since not all
observables behave in the same way \protect\cite{Lacks1994}, a general statement is not yet
possible. In fact it is conceivable, though unlikely, that specific
quantities might deteriorate so slowly with increasing $T$ as to remain
usable up to $T_m$. There are {\it a posteriori} hints that this behaviour
might hold for the crystallographic parameters.  In fact, in recent QHLD
calculations for naphthalene \protect\cite{Valle1995}, benzene \protect\cite{Valle1996b} and argon \protect\cite{Valle1996a} we obtained structural results in acceptable agreement with the
experiments in the whole range of existence of the solids. Since close to
$T_m$ the amplitude of the atomic vibrations are large and the harmonic
approximation is not expected to be accurate, these results require further
investigation. To check whether the good results for the crystallographic
structure near $T_m$ were a genuine consequence of the method, we have
performed QHLD and MD calculations for the structures of the Ne, Ar, Kr and
Xe crystals as a function of $p$ and $T$.

Following Lacks and Rutledge \protect\cite{Lacks1994}, we started this project \protect\cite{Valle1996a}
by using Verlet's \protect\cite{Verlet1967} LJ parameters for argon. These parameters are
known to give good agreement with the experimental properties of solid,
liquid and gaseous argon if many-body interactions are neglected \protect\cite{Verlet1967,Barker1977}. With this effective two-body potential we obtained \protect\cite{Valle1996a} excellent QHLD results for the density $\rho$, at least for very
low $T$.

Given the excellent results for Ar, we expected similarly good behaviour for
the heavier rare gases, Kr and Xe. We tried more than 20 different LJ models
from standard compilations \protect\cite{Allen1987,Hirschfelder1965,Batchelder1967,Crawford1977,Syassen1978,Kittel1996} and, much to our amazement, none of them gave
sensible results. All QHLD calculations with these models gave large
discrepancies with the experimental densities at low $T$: these are
precisely the conditions where QHLD was supposed to be accurate. Even more
puzzling was the fact that the {\it experimental} density did not follow any
regular trend with the atomic number once reduced with the literature LJ
parameters. Since the experimental data for Ne, Ar, Kr and Xe do follow a
regular trend when reduced with respect to the critical point data
($\rho_c$, $T_m$, $p_c$) \protect\cite{Hirschfelder1965}, we have concluded that the origin
of the discrepancies is the usage of inappropriate LJ parameters.

Perhaps one could have forecast these kinds of difficulties, since
most of the literature parameters \protect\cite{Allen1987,Hirschfelder1965,Batchelder1967,Crawford1977,Syassen1978,Kittel1996} are fitted to data on the
gas, rather than the solid. The gas data yield, in principle, the true
interaction potential between two isolated atoms, whereas for
the solid and liquid phases one needs effective two body models. Such
LJ models incorporate an average of the many body interactions and, 
therefore, can only be valid in a limited density range. We wish to
stress that we do not believe that the true interaction potential in
the rare gases follow a LJ curve, nor that the literature parameters
are ``wrong''. We aim to map the experimental behaviour of the rare
gas solids onto properly chosen LJ models. Since there are two
adjustable potential parameters, it is always possible to obtain exact
agreement between experiments and calculations for at least two
$p$,$T$ points, and approximate agreement for a limited range of
$p$,$T$ values. For our purposes we require agreement at low $p$ and
$T$ (the regime where QHLD is accurate), and we have therefore decided
to adjust the LJ models to low $p$, $T$ data. Different choices may be
appropriate for other purposes.

\section{Methods}

\subsection{Molecular Dynamics}

MD \protect\cite{Allen1987} is a simulation method in which the classical mechanical
equations of motion of the system are integrated numerically. The
method avoids any assumption of harmonicity of the lattice
vibrations, at the cost of neglecting all quantum effects. We use
the MD results as calibration data to investigate the convergence of
the QHLD results towards the classical limit, and to evaluate the
effects of the vibrational anharmonicity.

We performed a sequence of MD simulation runs for a specified number
of particles $N$, pressure $p$ and temperature $T$. The simulations
employed $N$=256 particles, in a cube with periodic boundary
conditions, and, using Andersen's isothermal-isobaric method \protect\cite{Andersen1980}, described an {\it fcc} crystal in contact with a heat
bath and subject to a hydrostatic pressure. The equations of motion
were integrated using the velocity Verlet algorithm \protect\cite{Verlet1967,Fox1984}. Since numerous MD studies of monoatomic LJ systems have been
reported \protect\cite{Allen1987,Verlet1967,Fox1984}, we omit other details of the
present simulation.

\subsection{Quasi Harmonic Lattice Dynamics}

In the QHLD method \protect\cite{Ludwig1967,Zharkov1971,Barron1974} the Gibbs
free energy of the crystal is  computed as the free energy of
an ensemble of phonons of frequency $\nu_i$:

$$ G(p,T) = \Phi_0 + pV + \sum_i \frac{h\nu_i}{2} + k_B T ~ \sum_i \ln\left[
1 - \exp\left(-\frac{h\nu_i}{k_BT}\right) \right]. \eqno(1) $$

\noindent
Here $\Phi_0$ is the total potential energy of the crystal in its average
structure (the electronic ground state energy), $pV$ is the pressure-volume
term, $\sum h\nu_i/2$ is the zero-point energy, and the last term is the
entropic contribution. The structure as a function of $p$ and $T$ is
determined by minimizing $G$ with respect to the unit cell volume. Further
details of the QHLD method appear in Refs. \protect\cite{Valle1995,Valle1996b,Valle1996a}.

For comparison with the MD calculations, where quantum effects are
neglected, it is appropriate to consider the classical limit ({\it i.e.} $h
\to 0$) of eq. (1) \protect\cite{Zharkov1971}:

$$ G_{\rm classic}(p,T) = \Phi_0 + pV +
 k_B T ~ \sum_i \ln\left(\frac{h\nu_i}{k_BT}\right). \eqno(2) $$

\noindent
It should be noticed that, although eq. (2) still depends on Planck's
constant $h$, its derivatives, and thus the equilibrium crystal structure,
do not. The difference between $G_{\rm classic}$ and $G$ equals the zero
point energy at 0 K, and, as appropriate for a classical limit, tends to
disappear at very high $T$ ($k_BT \gg h\nu_i$).

If all vibrational effects are neglected (the static lattice
approximation), both $G$ and $G_{\rm classic}$ reduce to the purely
mechanical part of the Gibbs free energy, $G_{\rm mech}(p) = \Phi_0 +
pV$, which is the free energy in the limit of a large atomic mass. An
energy minimization (EM) calculation for $G_{\rm mech}$ avoids the
determination of the phonon frequencies and is therefore less time
consuming than the complete calculation.

\section{Calculations}

We have used MD and QHLD to compute, as a function of $p$ and $T$, the
structure of a monoatomic {\it fcc} lattice with pairwise additive LJ
interactions, $\Phi(r) = 4 \epsilon [ (\sigma/r)^{12} - (\sigma/r)^{6} ]$. To
compare different atomic species we use non-dimensional reduced units, {\it
e.g.}  temperature $T^*=T/\epsilon$, pressure $p^*=p\sigma^3/\epsilon$ and
density $\rho^*=\rho \sigma^3/m$; here $m$ is the atomic mass. This is the
standard approach in LJ calculations for monoatomic systems \protect\cite{Allen1987,Verlet1967,Fox1984}. When expressed in reduced units, the MD, $G_{\rm classic}$ and
$G_{\rm mech}$ results turn out to be independent of the atomic species,
because the classical mechanics system has no energy, length and mass scales
other than $\epsilon$, $\sigma$ and $m$. On the other hand, the experiments
and the QHLD results with the quantum free energy $G$ (eq. 1) depend on
Planck's constant $h$, which in reduced units reads
$h^*=h/\sqrt{m\sigma^2\epsilon}$, and are therefore dependent on the atomic
number $Z$ \protect\cite{Pathria1991}. Because $h^*$ tends to zero for large $Z$ (as $m$,
$\sigma$ and $\epsilon$ all increase with $Z$), the reduced data should
monotonically converge towards a classical limit in the sequence Ne, Ar, Xe,
Kr.

\section{Results}
\subsection{Argon density as a function of pressure}

In the calculations for solid Ar we have used Verlet's \protect\cite{Verlet1967} LJ
parameters (table \ref{t:table}). We report in fig. 1 the results for the argon density
$\rho$ as a function of pressure $p$. The measured $\rho$ {\it vs} $p$ at 4,
77 and 293 K \protect\cite{Anderson1975,Grimsditch1986} are compared to the corresponding MD
and QHLD results. The main body of the figure contains the data for moderate
pressures, $p\le2$ GPa, which we discuss first. At 4 K, where the
anharmonicity is negligible, QHLD reproduces the low $p$ measurements almost
perfectly, while MD, which neglects the zero-point expansion on the lattice,
over-estimates the experimental density. At higher $T$, where anharmonicity
is large but quantum effects are less important, the opposite situation holds
and MD is found to be more accurate than QHLD. The difference between MD and
QHLD is only noticeable at very low $p$ and tend to vanish for increasing
$p$.

The good results of the LJ model for $p\le2$ GPa do not extend to
higher pressures, as shown by the inset of fig. 1 for the results up
to 30 GPa.  Since in this $p$ range MD and QHLD are barely
distinguishable and both disagree with the experiments \protect\cite{Grimsditch1986},
the failure of the calculations must be due to deficiencies of the
potential, rather than to the calculation method. After analyzing
their high $p$ data, Grimsditch {\etal} \protect\cite{Grimsditch1986} have stated
that for very dense solid argon there is no way to reproduce the
experimental results with a pair potential and that many-body
potentials have to be incorporated. Even though it is not yet ruled out
that an effective pair potential with a different functional form
could work, it is clear that the Verlet's LJ model \protect\cite{Verlet1967} is not
usable beyond 2 GPa.

\subsection{Dependence of the density on the atomic species}

As discussed in the introduction, none of the literature LJ models for
Ne, Kr and Xe that we have tested \protect\cite{Allen1987,Hirschfelder1965,Batchelder1967,Crawford1977,Syassen1978,Kittel1996} was found appropriate for our
purposes. We have therefore developed new LJ models by fitting the
QHLD calculations to the density measurements as a function of $p$ at
the lowest available $T$ and in a range of moderate pressures
(0-2 GPa). This is the regime where the harmonic approximation is
valid and, as shown by the results for Ar, effective two-body
LJ models are usable. The optimal LJ parameters, which are
reported in table \ref{t:table}, have values in the ranges of the typical
literature models. The Ne, Kr and Xe data \protect\cite{Anderson1975,Anderson1973} used in the
fit are compared in fig. 2 with the QHLD results with the optimal LJ
models. The Ar measurements and calculations at 4 K already shown in
fig. 1 are also reproduced in fig. 2, together with the results of a
EM calculation  for $G_{\rm mech}$ (static lattice approximation),
which in reduced units is independent of the atomic species.

The good agreement between experiments and calculations in fig. 2 has no
particular significance, as it merely indicates that the fits are good. Of
greater importance is the fact that the experimental data, when reduced
with this particular set of $\epsilon$ and $\rho$ values, follow a
sensible trend with the atomic mass $m$. The reduced density $\rho^*$ is
lower for Ne, which has the lighter atom and therefore the largest
zero-point effects, and then increases with $m$. As the classical limit is
approached in the Ne, Ar, Kr, Xe sequence, the experimental $\rho^*$ tends
to a limiting curve, which appears to be very close to the static lattice
calculations for infinite $m$. The $G_{\rm mech}$ results, which can be
obtained very economically, can be an useful approximation if one wants to
estimate the effects of pressure without all the complications of a
complete QHLD calculation.

To clarify the role of the quantum effects on the density, we have
computed $\rho^*(p,T)$ at zero $p$ and $T$ as a function of $m$,
considered as a continuously varying parameter. The resulting $\rho^*$
is displayed as a function of $h^*=h/\sqrt{m\sigma^2\epsilon}$~ in the
inset of fig. 2, together with the experimental data at the lowest
available $T$ and $p$. These data constitute, in effect, a vertical
cross section of fig. 2 along the $p=0$ line. Since the $\rho^*(0,0)$
{\it vs} $h^*$ curve does not depend on any physical dimension, all
the experimental data at $p\approx0$, $T\approx0$, if properly
reduced, should lie close to the curve. This is what happens with the
LJ parameters of table \ref{t:table}, but not with the other LJ models that we
have tried \protect\cite{Allen1987,Hirschfelder1965,Batchelder1967,Crawford1977,Syassen1978,Kittel1996}. This discrepancy is one of the problems
with the literature parameters mentioned in the introduction.

The $\rho^*$ {\it vs} $h^*$ curve, which is found to deviate
very little from a straight line, exhibits the expected trend towards
a limiting density with increasing atomic number. The $h^*=0$ limit
(infinite $m$) is $\rho^*=1.092$, the correct minimum energy
density for an {\it fcc} lattice of LJ particles \protect\cite{Kittel1996}.
For increasing $h^*$ (decreasing $m$), the zero-point effects
become progressively more important and the lattice expands.

\subsection{Temperature dependence of the density}

We report in fig. 3 the densities as a function of $T$. The
reduced experimental data for Xe, Kr, Ar and Ne at atmospheric
pressure are compared to the MD density (which in reduced units does
not depend on the atomic species) and to the QHLD densities.
With increasing $T$, quantum effects become less important and the
experimental density difference among the various species
decreases. The density differences, however, are still large when the
melting temperature $T_m^{\rm exp}$ is reached. This observation indicates that,
especially for neon, quantum effects are still significant even at
melting.

Since it ignores all quantum effects, MD cannot distinguish between
the different atomic species and over-estimates the density at $T=0$,
though only slightly for the heavier atoms. For very small $T$, MD
predicts a thermal expansion linear in $T$, whereas experimentally
the density is initially independent of $T$. This is a non-classical
zero-point effect, which is properly reproduced by the QHLD
calculations. For large $T$, where the anharmonicity is large
but quantum effects decrease, the MD results tend to agree with the
experiments better than QHLD. As expected, the agreement is much
better for Ar, Kr and Xe than for Ne, which remains a quantum
particle even close to the melting point.

QHLD agrees with the experimental densities near to $T=0$ (for Ne, Kr and
Xe, by construction). As $T$ is raised, QHLD predicts a thermal expansion
larger than that experimentally observed, because of the neglect of the
vibrational anharmonicity \protect\cite{Valle1995}. The discrepancy continues to
increase, until we encounter a limiting temperature $T_m$, different for
each atomic species, above which the QHLD calculations fails because no
minimum of $G$ can be found. The QHLD curves in figure 3 are truncated at
$T_m$, which is found to be very close to the experimental melting
temperature. Further discussion of this point is postponed to the next
section.

The results shown in fig. 3 confirm those of fig. 1 and indicate that QHLD
and MD are essentially complementary methods, since QHLD is accurate at
low $T$, while MD is better at high $T$. We have found that QHLD
calculations with $G_{\rm classic}$ provide an useful connection between
the two methods, as shown for Ar in fig. 4, where we compare the QHLD, MD,
and $G_{\rm classic}$ densities, $\rho_{\rm QHLD}$, $\rho_{\rm MD}$ and
$\rho_{\rm classic}$. At 0 K the $\rho_{\rm classic}$ and $\rho_{\rm MD}$
results coincide, since both structures correspond to the absolute minimum
of the potential energy. As $T$ is raised, $\rho_{\rm classic}$
progressively diverges from $\rho_{\rm MD}$, because anharmonicity is
increasing, and converges towards $\rho_{\rm QHLD}$, as the quantum
effects tend to decrease. The difference between $\rho_{\rm MD}$ and
$\rho_{\rm classic}$ is solely due to anharmonicity, whereas the
difference between $\rho_{\rm QHLD}$ and $\rho_{\rm classic}$ is only due
to quantum effects. Therefore one can estimate the anharmonic corrections
as $\rho_{\rm MD}-\rho_{\rm classic}$, and the quantum corrections as
$\rho_{\rm QHLD}-\rho_{\rm classic}$. Since interactions between
anharmonic and quantum effects are ignored, these are first order
estimates of the corrections.

By adding the anharmonic corrections to $\rho_{\rm QHLD}$, or, equivalently
the quantum corrections to $\rho_{\rm MD}$, one get a ``corrected''
density, $\rho_{\rm corr}=\rho_{\rm QHLD}+\rho_{\rm MD}-\rho_{\rm classic}$
(also shown in fig. 4). We find that the corrected densities agree with the
experiments better than either $\rho_{\rm QHLD}$ or $\rho_{\rm MD}$, with
deviations around 0.7 \% on the average (2 \% in the worst case) in the
whole range of stability of solid Ne, Ar, Kr and Xe. The anharmonic
corrections are usually small, 1.6 \% on the average and 3.5 \% in the
worst case, and do not change very much with the atomic species. The
quantum corrections are largest for Ne at 0 K (11 \%) and, as expected,
decrease in the Ar, Kr, Xe sequence (3.4, 1.7 and 1.3 \%, respectively). By
comparison, the deviations between experimental and calculated densities
for Van Der Waals ({\it i.e.} molecular) crystals are usually found to be
in the 1--5 \% range \protect\cite{Pertsin1987,Valle1995,Valle1996b}. Our results confirm that this
is indeed the typical accuracy of the MD and QHLD calculations themselves.

\subsection{Thermal expansion mechanism and melting}

Since the curvature of the interatomic potentials decreases with increasing
distances, lattice expansion leads to smaller phonon frequencies and,
consequently, to lower zero-point and entropic contributions to the free
energy $G$ (eq. 1). The thermal expansion of the lattice is driven by the
competition between the potential energy term $\Phi_0$, which favours
structures close to mechanical equilibrium, and the entropic term, which
favours expanded structures with smaller frequencies, and which becomes
progressively more important as $T$ increases. Fig. 5 shows the Gibbs free
energy $G(p,T)$, computed as a function of the density $\rho$, for argon at
$p=0$ and for several values of $T$. At $T \approx 0$, where the only
difference between $\Phi_0$ and $G$ is the zero-point energy, the
equilibrium density is slightly lower than that yielding the minimum of
$\Phi_0$. The equilibrium density decreases even further with increasing
$T$, since the entropic effects increase. Eventually a critical temperature
$T_m$ is reached above which the minimum of $G$ vanishes altogether: the
QHLD model become unstable for $T>T_m$. To clarify this behaviour we show
in fig. 5 the locus of the minima of $G$, which, by definition, represents
$G$ {\it vs} $\rho$ along the curve of thermodynamic equilibrium. We have
computed this curve as a parametric function of $T$, $G(0,T)$ {\it vs}
$\rho(0,T)$, following $T$ from 0 to the disappearance of the minimum,
which for Ar occurs at $T_m^*=0.651$ ($T_m=78$ K).

The idea that melting is connected to an intrinsic instability of the solid
goes back to Herzfeld and Goeppert Mayer \protect\cite{Herzfeld1934} and has been
discussed many times \protect\cite{Boyer1979,Boyer1980,Boyer1985,Born1939,Tallon1979,Ross1986}. As
convincingly argued by Ross and Wolf \protect\cite{Ross1986}, the various instability
models consider the solid phase only, and therefore cannot describe genuine
thermodynamic transitions between solid and liquid phases of equal Gibbs
free energy. A correct prediction of the $p$, $T$ melting curve requires an
accurate calculation of the free energy in the two phases. However, the
stability criteria \protect\cite{Boyer1979,Boyer1980,Boyer1985} are still useful: loss of stability,
although not necessarily exactly coincident with melting, cannot be too
distant from it \protect\cite{Valle1996b}.

A formal characterization of the instability $T_m$ is obtained \protect\cite{Herzfeld1934,Boyer1979,Boyer1980} by examining directly the conditions for the
existence and stability of a local minimum of $G$: $\partial G/\partial
V=0$ and $\partial^2 G/\partial V^2>0$. Here we are considering stability
with respect to small volume fluctuations away from the volume $V(p,T)$ of
thermodynamic equilibrium, {\it i.e.} we consider $p$, $T$ and $V$ as
independently varying variables \protect\cite{Sewell1987}. By expressing $G$ in terms of
the Helmotz free energy $F$, $G(p,T)=pV+F(V,T)$, the equilibrium and
stability conditions reduce to $p=-\left(\partial F/\partial V\right)_T$
(the state equations) and $\left(\partial^2 F/\partial V^2\right)_T =
\beta_T/V > 0$, where $\beta_T=-V\left(\partial p/\partial V\right)_T$ is
the isothermal bulk modulus. It is the vanishing bulk modulus or,
equivalently, the divergence of the isothermal compressibility $k_T =
1/\beta_T = -\left(\partial V/\partial p\right)_T/V$, which is the origin
of the instability at $T_m$ \protect\cite{Herzfeld1934,Boyer1979,Boyer1980}. This description is
entirely consistent with that observed in our QHLD computations.

In these computations, where $G$ is numerically minimized by varying the
molar volume $V$ in a sequence of finite steps, we find that for $T<T_m$
the volume converges to an equilibrium value, whereas for $T>T_m$ the
calculation fails because no minimum is found, as shown by fig. 5.  The
QHLD instability temperatures $T_m$ are compared in table \ref{t:table}
with the experimental melting temperatures $T_m^{\rm exp}$ of the rare
gases. In fig. 3 the experimental points end near $T_m^{\rm exp}$, whereas
the QHLD curves are truncated at $T_m$.  Notwithstanding the shortcomings
due to the use of a solid-only melting criterion \protect\cite{Ross1986}, and to the
neglect of the anharmonic contributions to the free energy \protect\cite{Lacks1994}, the
computed $T_m$ follow quite closely the experimental data.  As shown by
fig. 3, the QHLD calculations account for the quantum effects on melting,
since they reproduce the observed decrease of the reduced $T_m^{\rm exp}$
for the lighter atoms. This is a purely quantum effect, as, according to
the classical law of corresponding states, the reduced melting temperature
of classical LJ particles should be independent of the atomic species \protect\cite{Pathria1991}.

\section{Conclusions}

We have computed the density of solid Ne, Ar, Xe, and Kr through QHLD,
MD, and EM methods. In this way we have controlled the classical (MD),
harmonic (QHLD) and static lattice (EM) approximations, {\it i.e.}
estimated rather precisely their effects. Anharmonic effects are
negligible at low temperatures and even close to the melting point
they account only for a few percent of the density. Quantum effects are
only important for very light atoms (Ne) and tend to decrease with
increasing temperatures. Motional effects, which are usually large,
decrease for large atomic masses, low temperatures and
high pressures.

We find that QHLD can be used to compute the structural parameters, with
slowly degrading accuracy, for all temperatures up to that where the
model becomes unstable. This breakdown temperature turns out to be a fair
estimate of the melting point. The QHLD accuracy for the structural
parameters is much better than that for the free energy. This result is not
an artifact but a genuine property of the QHLD method, though
we do not have yet a fully satisfactory explanation for it.

QHLD and MD are found to be complementary, rather than competing, methods.
QHLD, where quantum effects are accounted for, but vibrational
anharmonicity is neglected, is the better method at low temperatures. MD,
which ignores all quantum effects but correctly handles large amplitude
vibrations, is appropriate for solids close to the melting point and for
fluids. The differences between MD and QHLD, in the range where both are
valid, are small. Since QHLD is much more efficient than MD, we think that
a QHLD computation should be one of the first steps in testing any proposed
potential model, even for problems where disorder or large amplitude motion
will eventually require MD or MC (Monte Carlo) techniques. If one wants to
study the effects of pressure, rather than temperature, even a $G_{\rm
mech}$ energy minimization (EM) can be a useful starting point. The EM method,
which neglects both quantum and anharmonic effects, is quite accurate for
heavy molecules at low temperatures.

For a given problem it may be important to choose the fastest method
among those applicable, as the speed differences are very large. For
Ar on a fast RS/6000 workstation we needed 0.2 sec for a EM
calculation, 20 sec for a QHLD optimization, and 75 sec for a MD
simulation with 256 atoms and 1000 timesteps, the minimum for barely
acceptable equilibration and statistics. Though these times obviously
depend on the specific problem, their ratios are expected to be quite
typical.

It should be noticed that for Ne, Kr and Xe we have fitted the
potential models solely to the experimental density as a function of
$p$ at very low $T$. Though no $T$ dependent data has been used in the
fits, the models reproduce well the density as a function of $T$,
within the accuracy of the calculation methods. These results provide
some justification for the procedure of neglecting thermal effects
while fitting a potential model. This approach, which is usually
adopted because it is very convenient, is found to lead to acceptable
results.

Our work shows that one can efficiently and accurately predict the
structure of a solid phase, through the whole range of its
thermodynamic stability, by using a suitable combination of EM, QHLD
and MD methods.

\acknowledgments
Work done with funds from the University of Bologna (``Finanziamento
speciale alle strutture''). We also thank MURST and CNR for further
financial support.

\begin{table}[ht]
\caption{Data for the rare gases. The table lists LJ potential parameters
$\epsilon$ and $\sigma$, mass $m$, reduced Planck constant
$h^*=h/\protect\sqrt{m\sigma^2\epsilon}$, and experimental \protect\cite{Crawford1977} and
computed melting temperatures $T_m^{\rm exp}$ and $T_m$.
LJ parameters for Ar are from Ref. \protect\cite{Verlet1967}; those for Ne, Kr and Xe 
are from our fits.}

\begin{tabular}{rrrrrrr}
   & $\epsilon$~~ & $\sigma$~ & $m$~~~ & $h^*$~ & 
   $T_m^{\rm exp}$ & $T_m$~ \\
   &  (K) & (\AA) & (amu) & & (K)~ & (K)~ \\
\hline
Ne &  38.5 & 2.786 &  20.183   & 0.563 &  24.553 &  22 \\
Ar & 119.8 & 3.405 &  39.948   & 0.186 &  83.806 &  78 \\
Kr & 159.9 & 3.639 &  83.8~~~~ & 0.104 & 115.763 & 105 \\
Xe & 220.9 & 3.962 & 131.3~~~~ & 0.065 & 161.391 & 145 \\
\end{tabular}
\label{t:table}
\end{table}

\begin{figure}
\caption{Argon density $\rho$ (g/cm$^3$) {\it vs} pressure $p$ (GPa).
Symbols: experiments \protect\cite{Anderson1975,Grimsditch1986} at 4 K (circles), 77 K
(triangles) and 293 K (squares). Curves: calculations with MD (broken
lines) and QHLD (solid lines). The data at 293 K are also reproduced
in the inset, for pressures up to 30 GPa \protect\cite{Grimsditch1986}. The MD and
QHLD curves are not distinguishable at the scale of the inset.}
\end{figure}

\begin{figure}
\caption{Reduced density $\rho^*=\rho \sigma^3/m$ {\it vs} 
reduced pressure $p^*=p \sigma^3/\epsilon$
for the rare gases at low temperatures ($T \approx 0$).
Experiments at 4 K for
Ne \protect\cite{Anderson1973},
Ar,
Kr and
Xe \protect\cite{Anderson1975}:
symbols as indicated in the figure. QHLD calculations with $G$ and $G_{\rm
mech}$: solid and broken lines, respectively. The inset displays $\rho^*$
{\it vs} $h^*=h/\protect\sqrt{m\sigma^2\epsilon}$ at $p=0$: QHLD calculations at
$T=0$ (solid line) and experiments \protect\cite{Korpiun1977} on
Ne at 3 K,
Ar and Kr at 4 K, and
Xe at 5 K (symbols).}
\end{figure}

\begin{figure}
\caption{Reduced density $\rho^*$ {\it vs} reduced temperature
$T^*=T/\epsilon$ for the rare gases at atmospheric pressure.
Experiments on
Ne \protect\cite{Batchelder1967,Korpiun1977},
Ar \protect\cite{Korpiun1977,Peterson1966,Schwalbe1980},
Kr \protect\cite{Korpiun1977,Loose1968} and
Xe \protect\cite{Korpiun1977,Sears1962,Eatwell1961}: symbols as indicated in the figure.
QHLD and MD results: solid and broken lines, respectively.}
\end{figure}

\begin{figure}
\caption{Reduced density $\rho^*$ {\it vs} reduced temperature $T^*$ for Ar
at atmospheric pressure. Experiments \protect\cite{Korpiun1977,Peterson1966,Schwalbe1980}:
circles; MD calculations, ``corrected'' density, QHLD results with $G$ and
with $G_{\rm classic}$: lines as indicated in the figure.}
\end{figure}

\begin{figure}
\caption{The solid curves represent the reduced QHLD Gibbs free energy
$G^*(p^*=0,T^*) = G/\epsilon$ of Ar, computed as a function of the
reduced density $\rho^*$, for several values of $T^*$ (namely $T^*$ =
0.1, 0.2, 0.3 $\ldots$ 0.7, 0.8), as indicated near each curve. The
broken line represents the locus of the minima of $G^*$, as discussed
in the text, followed from $T^*=0$ to the disappearance of the
minimum at $T^*=0.651$.}
\end{figure}

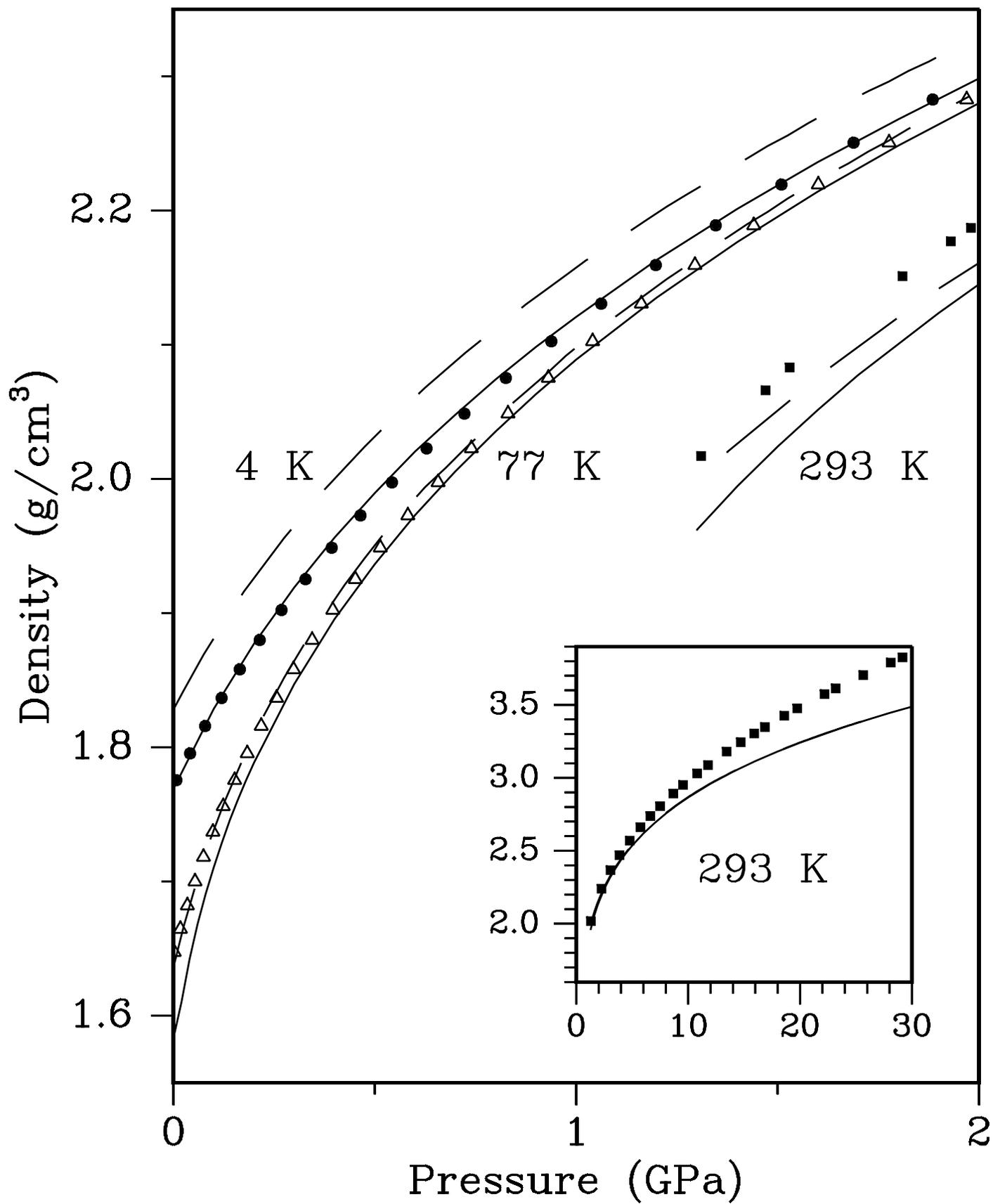
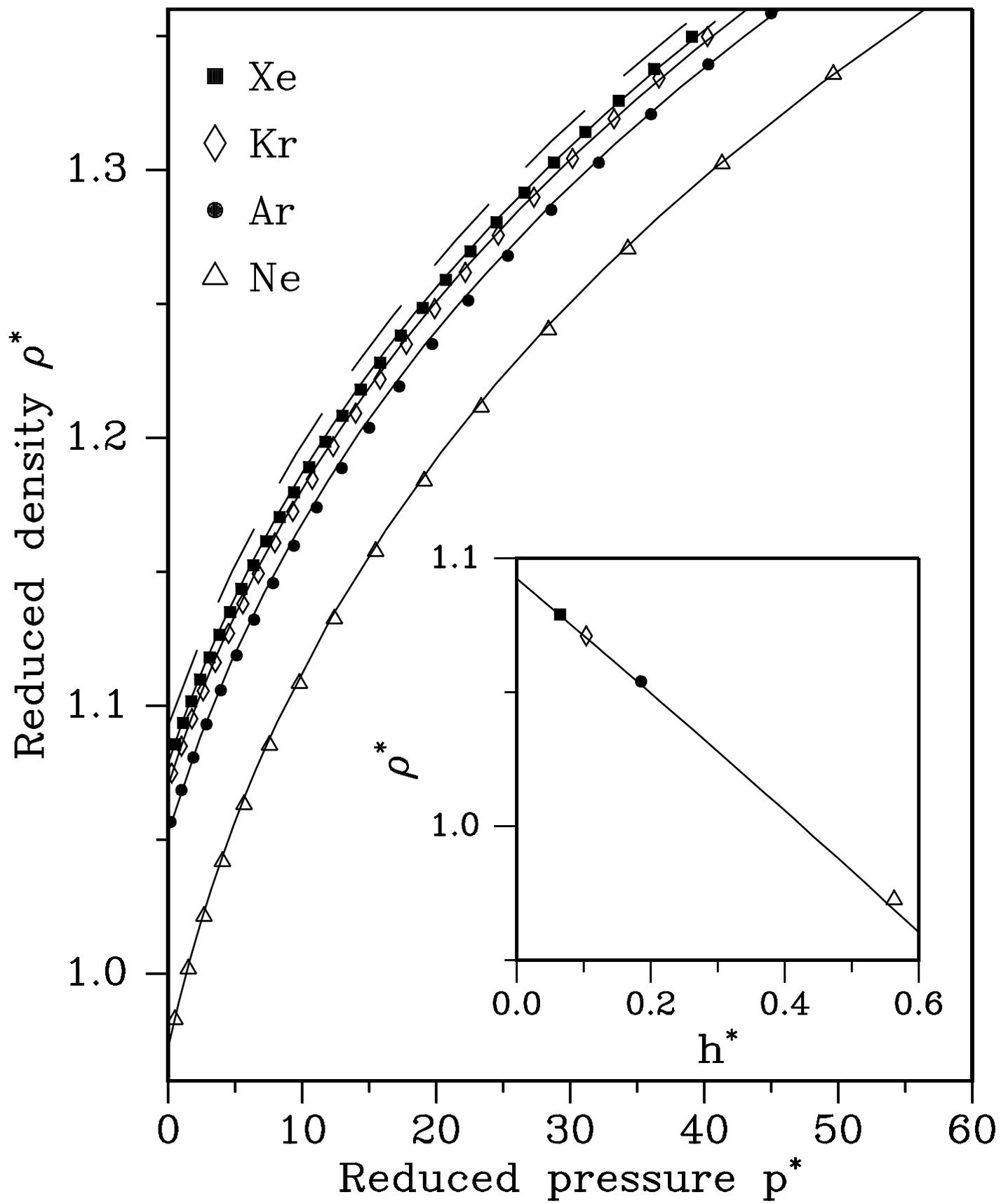
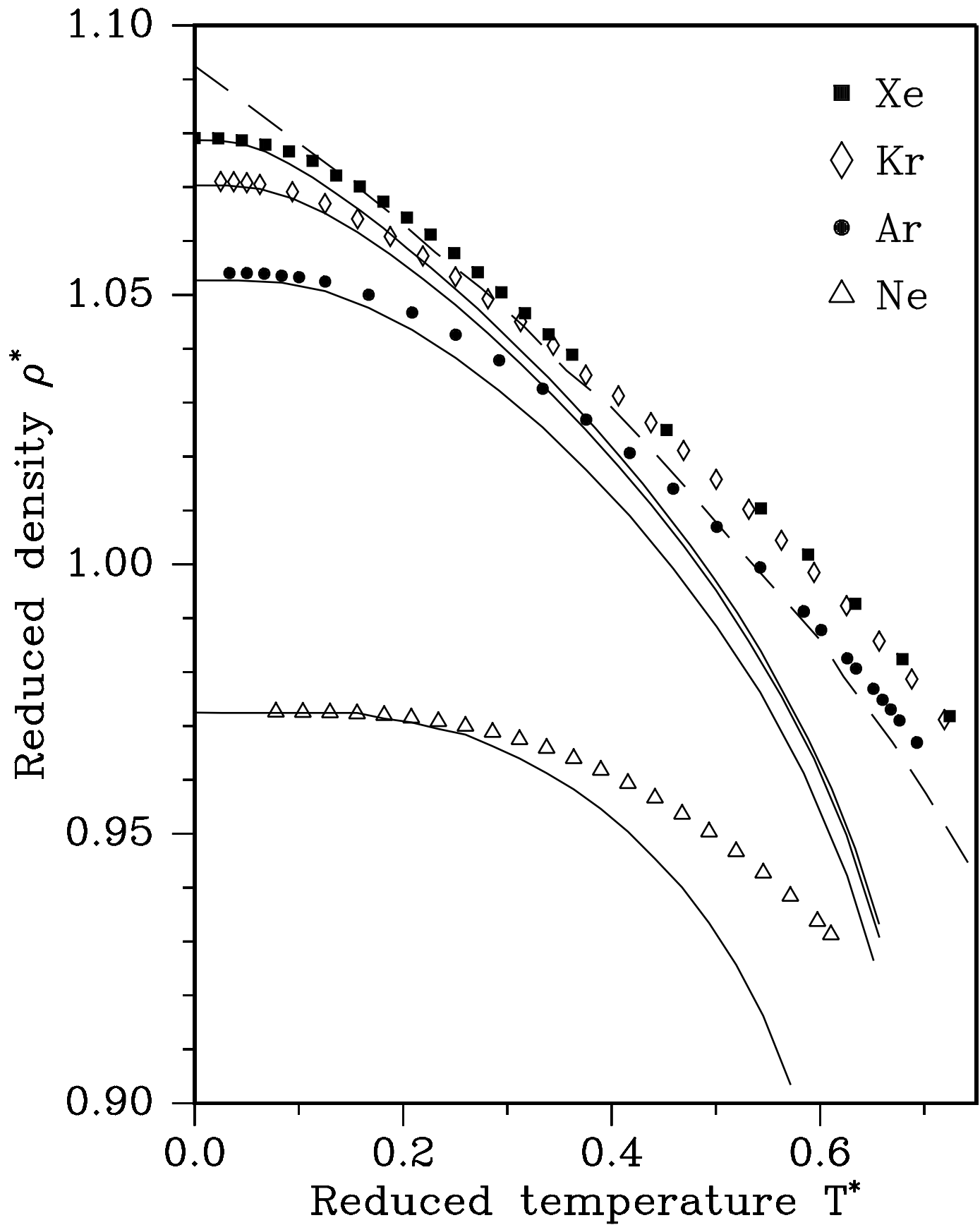
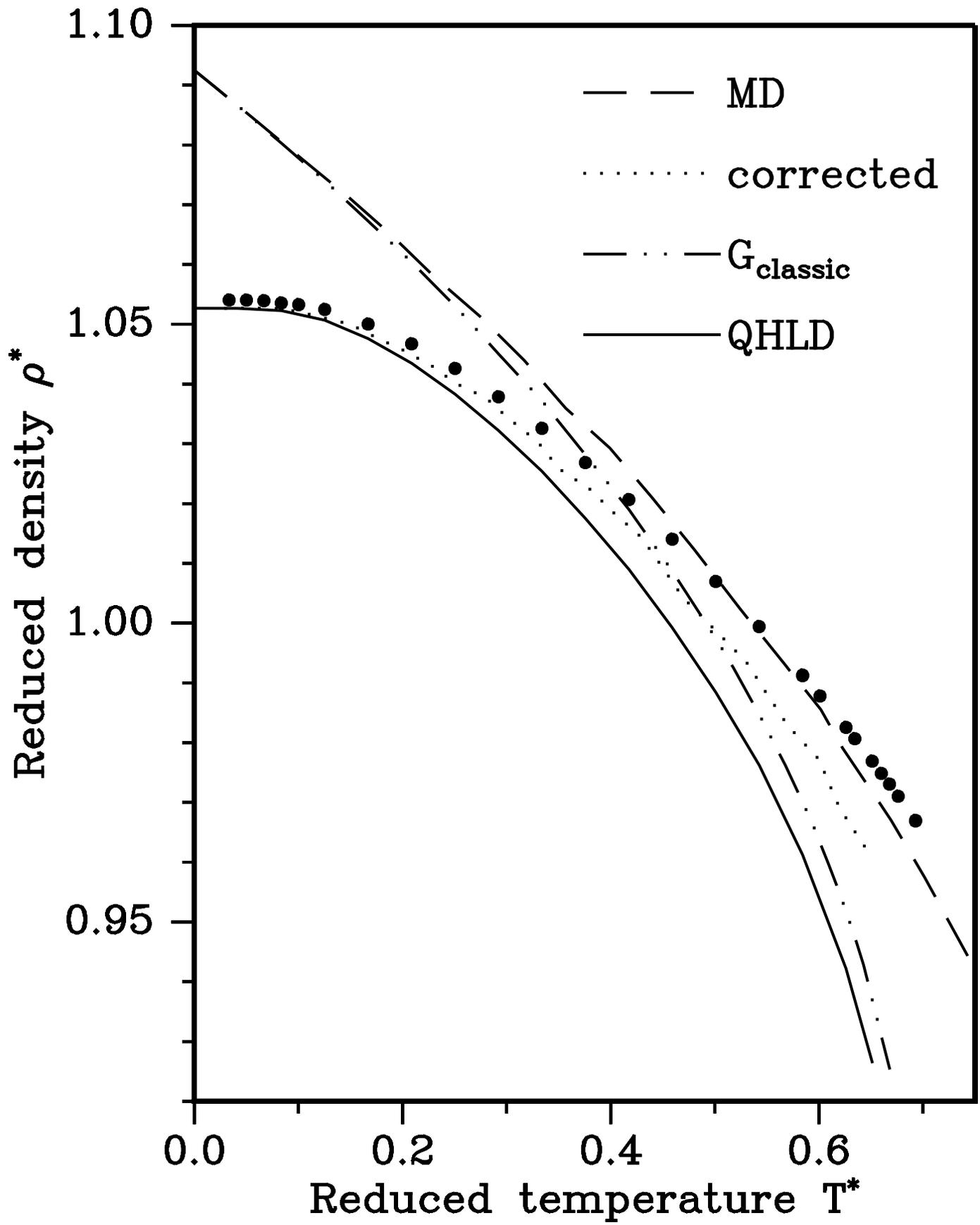
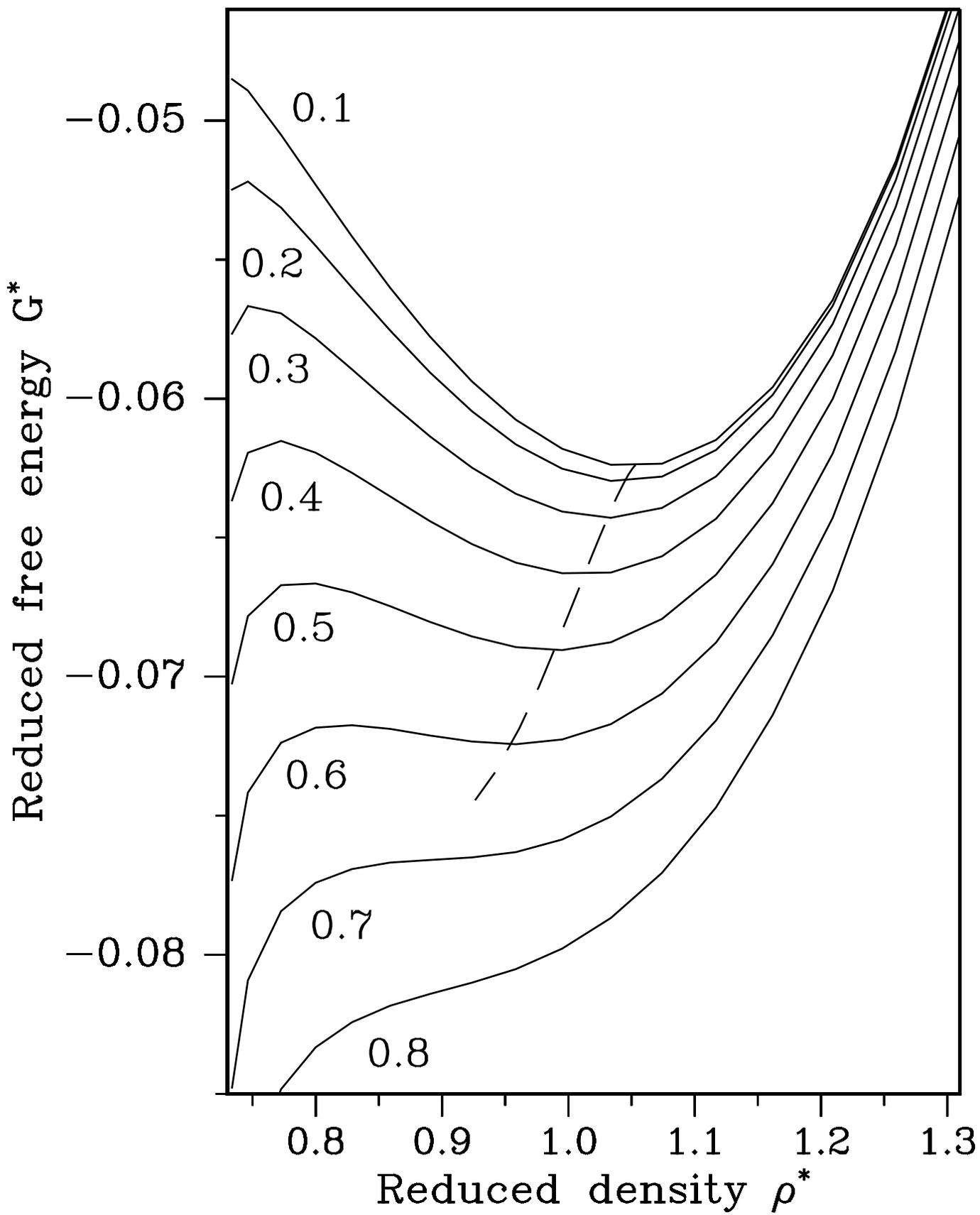
\end{document}